Surface plasmon excitation by a quantum oscillator

V.V.Lidsky

The P.N.Lebedev Physical Institute of RAS
119991 Moscow, Russia
vlidsky@sci.lebedev.ru

Surface waves in a thin uniform metal film are described in terms of quantum electrodynamics. The interaction of surface waves with a quantum oscillator is discussed in the dipole approximation. The increase in the spontaneous emission rate of the excited quantum oscillator, the so called Purcell factor, is evaluated to be as high as $10^5$.

Взаимодействие поверхностных плазмонов с квантовым излучателем

В.В.Лидский

Физический институт им. П. Н. Лебедева РАН
119991 Москва, Ленинский пр., 53
vlidsky@sci.lebedev.ru

А Н Н О Т А Ц И Я

Рассмотрены поверхностные волны в тонкой металлической пленке в рамках формализма квантовой электродинамики. Рассмотрено в дипольном приближении взаимодействие поверхностных плазмонов с квантовым излучателем, расположенным в непосредственной близости от поверхности пленки. Показано, что вероятность спонтанного перехода с излучением плазмона превосходит вероятность дипольного перехода в свободном пространстве, причем фактор увеличения вероятности может достигать значений $10^5$.

1. После появления в печати сообщений о запуске нанолазеров или SPAZER'ов /1,2/, использующих механизм возбуждения поверхностных волн молекулами красителя, возник вопрос о построении последовательной квантовой теории взаимодействия поверхностных волн с квантовым излучателем. Дело в том, что вероятность спонтанного перехода квантовой системы существенным образом зависит от структуры пространства, окружающего излучатель. Перселл заметил, что вероятность спонтанного перехода возрастает на несколько порядков, если вблизи излучателя находится микроскопическая частица /3/. В /4/ эффект Перселла был привлечен для объяснения наблюдаемого увеличения на 14 порядков сечения комбинационного рассеяния — явления гигантского комбинационного рассеяния (SERS) /5,6/. Теория SPAZER'а была предложена в недавно появившейся работе /7/, где показано, что и здесь учет эффекта Перселла позволяет в принципе преодолеть трудность, вызванную исключительно сильным затуханием поверхностных волн.

В нашей работе мы рассмотрим поверхностные волны, распространяющиеся по тонкой металлической пленке, окруженной диэлектрической средой. Затем мы разложим поле на элементарные моды и определим операторы вторичного квантования, с помощью которых поле поверхностной волны может быть описано с точки зрения квантовой электродинамики. Затем рассмотрим квантовый осциллятор, помещенный вблизи поверхности пленки и вычислим вероятность излучения этим осциллятором поверхностного плазмона. Полученную величину сравним с известной формулой для вероятности спонтанного излучения фотона в свободном пространстве и придем к квантово-механической формулировке эффекта Перселла.

2. Рассмотрим тонкую металлическую пленку в диэлектрической среде. Диэлектрическую проницаемость среды с обеих сторон пленки будем считать равной $\varepsilon_h$. Диэлектрическую проницаемость металла будем считать связанной с частотой формулой Друдэ: $\varepsilon_m = 1 - \omega_{pl}^2 / \omega^2$. Мнимая часть диэлектрической проницаемости в данной работе учитываться не будет. Магнитную проницаемость металлической пленки будем считать равной 1.

Введем декартову систему координат, так чтобы оси $x$, $y$ располагались в плоскости пленки. Ось $z$ направим перпендикулярно плоскости пленки, см. рис. 1. Две поверхности пленки имеют координаты: $z_b = 0$ и $z_h = d$.

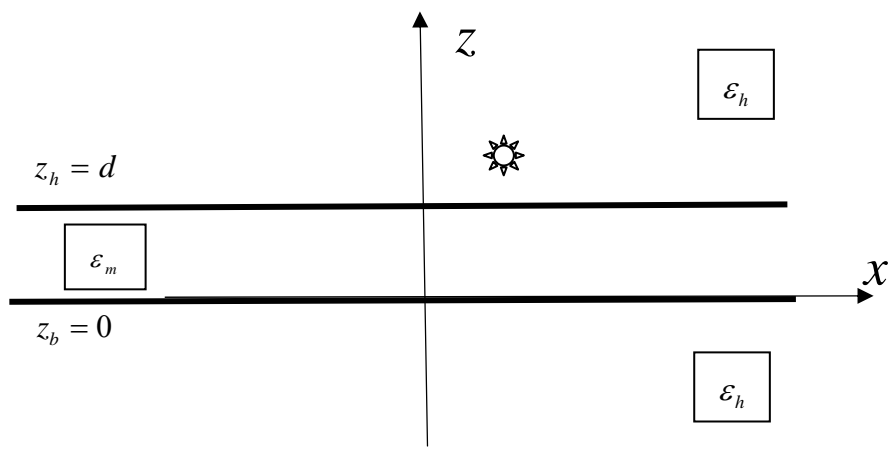

Рис.1.

На первом шаге наша задача определить собственные поверхностные волны такого пространства. Уравнения Максвелла при отсутствии внешних токов запишем в виде:

$$rot\ \vec{H}\ =\ \varepsilon \cdot \partial_t \vec{E} \qquad rot\ \vec{E}\ =\ -\partial_t \vec{H} \tag{2.1}$$

$$div\ \vec{H}\ =\ 0 \qquad div\ \vec{E}\ =\ 0 \tag{2.2}$$

Мы видим, что уравнения (2.2) выполнены, если выполнены (2.1).

Аналогично тому, как это делается в свободном пространстве /8/, выберем поверхность пленки $S$, ограниченную прямоугольником со сторонами $a$ и $b$. Разложим поля в ряды Фурье по координатам $x$ и $y$., а по времени не будем переходить к Фурье образам.

$$\vec{E} = \sum_k \left( \vec{E}_k(t,z) \cdot \exp(ik_x \cdot x + ik_y \cdot y) + \vec{E}_k^*(t,z) \cdot \exp(-ik_x \cdot x - ik_y \cdot y) \right)$$

$$\vec{H} = \sum_k \left( \vec{H}_k(t,z) \cdot \exp(ik_x \cdot x + ik_y \cdot y) + \vec{H}_k^*(t,z) \cdot \exp(-ik_x \cdot x - ik_y \cdot y) \right) \tag{2.3}$$

Величины $\vec{E}_k(t,z), \vec{H}_k(t,z)$ зависят от времени по закону

$$\vec{E}_k(t,z) \propto e^{-i\omega t} \tag{2.4}$$

причем частота $\omega$ определяется из известного характеристического уравнения:

$$\exp(p \cdot d) = \frac{(\varepsilon_m \cdot q - \varepsilon_h \cdot p)}{(\varepsilon_m \cdot q + \varepsilon_h \cdot p)} \tag{2.5}$$

В которое частота $\omega$ входит через следующие выражения:

$$q = \sqrt{k^2 - \varepsilon_h \cdot \omega^2} \qquad\qquad p = \sqrt{k^2 - \varepsilon_m \cdot \omega^2} \tag{2.6}$$

где

$$k = \sqrt{k_x^2 + k_y^2} \tag{2.7}$$

Для компонент векторов $\vec{E}_k(t,z), \vec{H}_k(t,z)$ справедливы следующие соотношения, вытекающие из уравнений Максвелла с нулевой правой частью:

$$ik_y H_{kz} - \partial_z H_{ky} = -i\omega\varepsilon \cdot E_{kx} \qquad\qquad ik_y E_{kz} - \partial_z E_{ky} = i\omega \cdot H_{kx}$$
$$\partial_z H_{kx} - ik_x H_{kz} = -i\omega\varepsilon \cdot E_{ky} \qquad\qquad \partial_z E_{kx} - ik_x E_{kz} = i\omega \cdot H_{ky} \qquad (2.8)$$
$$ik_x H_{ky} - ik_y H_{kx} = -i\omega\varepsilon \cdot E_{kz} \qquad\qquad ik_x E_{ky} - ik_y E_{kx} = i\omega \cdot H_{kz}$$

3. Введем в плоскости пленки новые координаты $\xi, \eta$ так, чтобы ось $\xi$ совпадала с направлением вектора $(k_x, k_y)$, а ось $\eta$ направим в плоскости пленки в перпендикулярном к оси $\xi$ направлении, при этом третья ось, ось $z$, (обозначение которой мы не меняем) сохраняет свое направление. Для компонент векторов напряженностей находим:

$$H_{k\xi} = \frac{1}{k}(k_x H_{kx} + k_y H_{ky}) \qquad\qquad H_{k\eta} = \frac{1}{k}(-k_y H_{kx} + k_x H_{ky})$$
$$E_{k\xi} = \frac{1}{k}(k_x E_{kx} + k_y E_{ky}) \qquad\qquad E_{k\eta} = \frac{1}{k}(-k_y E_{kx} + k_x E_{ky}) \qquad (3.1)$$

После такой замены неизвестных система (2.8) превращается в систему шести уравнений, которая распадается на две независимые тройки уравнений. Первая тройка уравнений:

$$\begin{cases} \partial_z H_{k\eta} - \varepsilon i\omega E_{k\xi} = 0 \\ k \cdot H_{k\eta} + \varepsilon\omega E_{kz} = 0 \\ \partial_z E_{k\xi} - ik \cdot E_{kz} - i\omega H_{k\eta} = 0 \end{cases} . \qquad (3.2)$$

Вторая тройка:

$$\begin{cases} \partial_z H_{k\xi} - ik \cdot H_{kz} + \varepsilon i\omega E_{k\eta} = 0 \\ \partial_z E_{k\eta} + i\omega H_{k\xi} = 0 \\ k \cdot E_{k\eta} - \omega H_{kz} = 0 \end{cases} . \qquad (3.3)$$

Несложно показать, что система (3.3) не имеет решений, соответствующих свободным поверхностным волнам. Рассмотрим решение системы (3.2). Преобразуя (3.2), находим уравнение, определяющее зависимость компоненты $E_{kz}$ поля от координаты $z$:

$$\partial_z^2 E_{kz} - (k^2 - \varepsilon\omega^2) \cdot E_{kz} = 0 \qquad (3.4)$$

Где $\varepsilon$ — диэлектрическая проницаемость в каждой из трех сред (см. рис. 1).

4. Будем обозначать индексами (b), (m), (h) физические величины, относящиеся к средам ниже пленки, внутри пленки и выше пленки соответственно. Тогда из (3.2) и (3.4) следует, что

$$H_{k\eta}^{(b)} = C_b \cdot \exp(q \cdot z)$$
$$H_{k\eta}^{(m)} = C_m \cdot \exp(p \cdot z) + C_r \cdot \exp(-p \cdot z) \qquad (4.1)$$
$$H_{k\eta}^{(h)} = C_h \cdot \exp(-q \cdot z)$$

Здесь $C_b, C_m, C_r, C_h$ — некоторые константы, соотношение между которыми определяются условиями на поверхностях пленки. Подставив (4.1) в первое из уравнений (3.2), получим

$$E_{k\xi}^{(b)} = \frac{1}{i\varepsilon_h\omega} \cdot C_b \cdot q \cdot \exp(q \cdot z)$$
$$E_{k\xi}^{(m)} = \frac{1}{i\varepsilon_m\omega} \cdot (p \cdot C_m \cdot \exp(p \cdot z) - p \cdot C_r \cdot \exp(-p \cdot z)) \qquad (4.2)$$
$$E_{k\xi}^{(h)} = -\frac{q}{i\varepsilon_h\omega} \cdot C_h \cdot \exp(-q \cdot z)$$

Требование непрерывности тангенциальных компонент $H_{k\eta}$ и $E_{k\xi}$ на каждой из поверхностей пленки приводит к системе уравнений:

$$C_b = C_m + C_r$$

$$C_h \cdot \exp(-q \cdot d) = C_m \cdot \exp(p \cdot d) + C_r \cdot \exp(-p \cdot d)$$

$$\frac{q}{\varepsilon_h} \cdot C_b = \frac{p}{\varepsilon_m} \cdot (C_m - C_r) \qquad (4.3)$$

$$-\frac{q_h}{\varepsilon_h} \cdot C_h \cdot \exp(-q_h \cdot d) = \frac{p}{\varepsilon_m} \cdot (C_m \cdot \exp(p \cdot d) - C_r \cdot \exp(-p \cdot d))$$

Система (4.3) имеет нетривиальные решения при выполнении условия (2.5). В этом случае с точностью до произвольного множителя $\gamma$ находим :

$$C_m = \gamma \cdot \frac{K+1}{2} \qquad\qquad C_r = -\gamma \cdot \frac{(K-1)}{2} \qquad C_h = -\gamma \cdot \exp(q \cdot d) \qquad C_b = \gamma \qquad (4.4)$$

где

$$K = \frac{\varepsilon_m \cdot q}{\varepsilon_h \cdot p} \qquad (4.5)$$

Таким образом, выражения (2.3) в каждой из трех сред можно переписать в виде:

$$\vec{E}_k^{(u)}(t,z) = \gamma \cdot \vec{e}_k^{(u)} \cdot \exp(-i\omega t) \cdot \exp(q^{(u)} z)$$

$$\vec{H}_k^{(u)}(t,z) = \gamma \cdot \vec{h}_k^{(u)} \cdot \exp(-i\omega t) \cdot \exp(q^{(u)} z) \qquad (4.6)$$

Индекс (u) означает одну из четырех волн: (h),(m),(r),(b).
При этом

$$q^{(h)} = -q \qquad\qquad q^{(m)} = p \qquad\qquad q^{(r)} = -p \qquad\qquad q^{(b)} = q \qquad (4.7)$$

Векторы $\vec{e}_k^{(u)}, \vec{h}_k^{(u)}$, определяющие поляризацию каждой из волн, имеют следующие отличные от ноля компоненты в координатах $\xi, \eta, z$ :

$$e_{k\xi}^{(h)} = -\frac{iq}{\varepsilon_h \cdot \omega} \cdot \exp(qd) \qquad e_{k\xi}^{(m)} = -\frac{ip}{\varepsilon_m \cdot \omega} \cdot \frac{K+1}{2} \qquad e_{k\xi}^{(r)} = -\frac{ip}{\varepsilon_m \cdot \omega} \cdot \frac{K-1}{2} \qquad e_{k\xi}^{(b)} = -\frac{iq}{\varepsilon_h \cdot \omega}$$

$$e_{kz}^{(h)} = \frac{k}{\varepsilon_h \cdot \omega} \cdot \exp(qd) \qquad e_{kz}^{(m)} = -\frac{k}{\varepsilon_m \cdot \omega} \cdot \frac{K+1}{2} \qquad e_{kz}^{(r)} = \frac{k}{\varepsilon_m \cdot \omega} \frac{K-1}{2} \qquad e_{kz}^{(b)} = -\frac{k}{\varepsilon_h \cdot \omega} \qquad (4.8)$$

$$h_{k\eta}^{(h)} = -\exp(qd) \qquad\qquad h_{k\eta}^{(m)} = \frac{K+1}{2} \qquad\qquad h_{k\eta}^{(r)} = -\frac{K-1}{2} \qquad\qquad h_{k\eta}^{(b)} = 1$$

Мы можем определить вектор-потенциал поля так, чтобы имели место соотношения:

$$\vec{H} = rot\vec{A} \qquad\qquad\qquad \vec{E} = -\partial_t \vec{A} - \nabla\Phi \qquad (4.9)$$

4-вектор потенциала $A^i$ можно определить выражением:

$$A_k^{(u)\mu} = \gamma \cdot a_k^{(u)\mu} \cdot \exp(q^{(u)} z - i\omega t) \qquad (4.10)$$

Причем $a_k^{(u)\mu}$ выберем так, чтобы он имел только две отличные от ноля компоненты:

$$a_k^{(h)3} = -\frac{i}{k} \cdot \exp(qd) \qquad a_k^{(m)3} = \frac{i}{k} \cdot \frac{K+1}{2} \qquad a_k^{(r)3} = -\frac{i}{k} \cdot \frac{K-1}{2} \qquad a_k^{(b)3} = \frac{i}{k}$$

$$a_k^{(h)0} = \frac{q}{k \cdot \varepsilon_h \omega} \exp(qd) \qquad a_k^{(m)0} = \frac{p}{k \cdot \varepsilon_m \omega} \cdot \frac{K+1}{2} \qquad a_k^{(h)0} = \frac{p}{k \cdot \varepsilon_m \omega} \cdot \frac{K-1}{2} \qquad a_k^{(b)0} = \frac{q}{k \cdot \varepsilon_h \omega} \qquad (4.11)$$

5. Обратимся теперь к вычислению энергии, переносимой каждой модой. Плотность энергии и поток энергии поля выражаются известными формулами:

$$w = \frac{1}{8\pi} \cdot (\varepsilon E^2 + H^2) \qquad\qquad\qquad \vec{S} = \frac{1}{4\pi} \cdot [\vec{E} \times \vec{H}] \qquad (5.1)$$

При этом, как легко убедиться из уравнений Максвелла (1) имеет место закон сохранения энергии:

$$\frac{d}{dt}\iiint_V w \cdot dV = -\oiint_\Sigma \vec{S} \cdot \vec{n} \cdot d\sigma \qquad (5.2)$$

Где $\Sigma$ -- поверхность, ограничивающая объем $V$. В случае поверхностной волны единственная отличная от ноля компонента вектора Пойнтинга — $S_\xi$. То есть вектор Пойнтинга коллинеарен волновому вектору поверхностной волны $\vec{k} = \{k_x, k_y, 0\}$.

$$S_\xi = \frac{1}{4\pi}\left(-E_{kz} \cdot H_{k\eta}^* - E_{kz}^* \cdot H_{k\eta}\right) = \frac{k}{2\pi\omega\varepsilon} \cdot |H_{k\eta}|^2 \qquad (5.3)$$

Любопытно, что вне металла поток энергии направлен вдоль $\vec{k}$, а внутри пленки – в противоположную сторону ($\varepsilon_m < 0$).

Для вычисления полной энергии, переносимой рассматриваемой волной в расчете на единичную площадь поверхности вычислим интеграл по объему из (5.2). При интегрировании по координатам $x, y$ произведений вида $\exp(ik_x x) \cdot \exp(ik_x' x)$ ненулевой вклад дадут только члены с $k'_x = -k_x$. После чего интегрирование по поверхности даст площадь $S$ рассматриваемой поверхности:

$$W = \frac{1}{8\pi}\iiint_V (\varepsilon E^2 + H^2) dV = \frac{S}{8\pi} \cdot \sum_{\vec{k}=-\infty}^{\infty} \int_{-\infty}^{\infty} dz(\varepsilon \vec{E}_k \vec{E}_k^* + \vec{H}_k \vec{H}_k^*) \qquad (5.4)$$

таким образом энергия распадается на сумму энергий, переносимых каждой модой:

$$W = \sum_{\vec{k}=-\infty}^{\infty} W_k \qquad (5.5)$$

где

$$W_k = \frac{S}{8\pi} \cdot \left(\int_{-\infty}^0 dz(\varepsilon\vec{E}_k\vec{E}_k^* + \vec{H}_k\vec{H}_k^*) + \int_0^d dz(\varepsilon\vec{E}_k\vec{E}_k^* + \vec{H}_k\vec{H}_k^*) + \int_d^\infty dz(\varepsilon\vec{E}_k\vec{E}_k^* + \vec{H}_k\vec{H}_k^*)\right) \qquad (5.6)$$

Вычислим последовательно интегралы из (5.6), используя (4.6) и (4.8). После преобразований получим выражение для энергии (см. Приложение 1):

$$W_k = \frac{S}{4\pi} \cdot \gamma \cdot \gamma^* \cdot \frac{k^2 \cdot (\varepsilon_h - \varepsilon_m)}{\varepsilon_h p^2 q}\left(1 + \frac{qd}{2}\left(\frac{(\varepsilon_m + \varepsilon_h)}{\varepsilon_h} - \frac{\varepsilon_m\omega^2}{k^2}\right)\right) \qquad (5.7)$$

Определим канонические переменные

$$q_k = (\gamma \cdot e^{-i\omega t} + \gamma^* \cdot e^{i\omega t})\frac{\sqrt{\Theta_k}}{\omega} \qquad\qquad p_k = -i(\gamma \cdot e^{-i\omega t} - \gamma^* \cdot e^{i\omega t})\sqrt{\Theta_k} \qquad (5.8)$$

где

$$\Theta_k = \frac{S}{8\pi} \cdot \frac{k^2 \cdot (\varepsilon_h - \varepsilon_m)}{\varepsilon_h p^2 q}\left(1 + \frac{qd}{2}\left(\frac{(\varepsilon_m + \varepsilon_h)}{\varepsilon_h} - \frac{\varepsilon_m\omega^2}{k^2}\right)\right) \qquad \textcolor{red}{(5.9)}$$

Очевидно, что

$$\dot{q}_k = p_k \qquad\qquad \dot{p}_k = -\omega^2 q_k \qquad (5.10)$$

Теперь энергия, колебаний на данной моде выражается через $p_k, q_k$:

$$W_k = \frac{1}{2}\left(\omega^2 \cdot q_k^2 + p_k^2\right) \qquad (5.11)$$

Функция Гамильтона системы поверхностных волн имеет вид:

$$H = \frac{1}{2}\sum_k \left( p_k^2 + \omega^2 \cdot q_k^2 \right)$$

(5.12)

6. При переходе к квантовой теории мы должны рассматривать канонические переменные $p_k, q_k$ как операторы с правилом коммутации:

$$\hat{p}_k \hat{q}_k - \hat{q}_k \hat{p}_k = -i\hbar$$

(6.1)

Теперь мы можем выразить через операторы $\hat{q}_k, \hat{p}_k$ компоненты напряженности и потенциала электромагнитного поля. Учитывая (5.8) из (4.6) и (4.10) находим:

$$\widehat{\vec{E}}_k^{(u)}(t,z) = \frac{\vec{e}_k^{(u)}}{2\sqrt{\Theta_k}} \cdot (i\hat{p}_k + \omega\hat{q}_k) \cdot \exp\!\left( q^{(u)} z \right)$$

$$\widehat{\vec{H}}_k^{(u)}(t,z) = \frac{\vec{h}_k^{(u)}}{2\sqrt{\Theta_k}} \cdot (i\hat{p}_k + \omega\hat{q}_k) \cdot \exp\!\left( q^{(u)} z \right)$$

(6.2)

$$\widehat{A}_k^{(u)\mu}(t,z) = \frac{a_k^{(u)\mu}}{2\sqrt{\Theta_k}} \cdot (i\hat{p}_k + \omega\hat{q}_k) \cdot \exp\!\left( q^{(u)} z \right)$$

(6.3)

где величины $q^{(u)}, \vec{e}_k^{(u)}, \vec{h}_k^{(u)}, a_k^{(u)i}$ определены в (4.7),(4.8), (4.11).
Гамильтониан системы плазмонов имеет вид вполне аналогичный (5.12):

$$\widehat{H} = \frac{1}{2}\sum_k \left( \hat{p}_k^2 + \omega_k^2 \cdot \hat{q}_k^2 \right)$$

(6.4)

7. Гамильтониан распадается на сумму гармонических осцилляторов. Собственные значения такого гамильтониана хорошо известны:

$$E_n = \left( n + \frac{1}{2} \right) \cdot \hbar \omega_k$$

(7.1)

А для квадратов модулей матричных элементов операторов $\hat{q}_k, \hat{p}_k$ переходов между собственными состояниями гамильтониана можно получить соотношения:

$$\left| \left\langle n_k \middle| \hat{q}_k \middle| n_k - 1 \right\rangle \right|^2 = \left| \left\langle n_k - 1 \middle| \hat{q}_k \middle| n_k \right\rangle \right|^2 = \frac{\hbar}{2\omega_k} \cdot n_k$$

$$\left| \left\langle n_k \middle| \hat{p}_k \middle| n_k - 1 \right\rangle \right|^2 = \left| \left\langle n_k - 1 \middle| \hat{p}_k \middle| n_k \right\rangle \right|^2 = \frac{\hbar \omega_k}{2} \cdot n_k$$

(7.2)

Ясно, что всегда можно, начав с первого, умножить собственные векторы на фазовый множитель так, чтобы матричные элементы оператора $\hat{q}_k$ были вещественны и положительны. В этой системе собственных векторов мы находим:

$$\left\langle n_k \middle| \hat{q}_k \middle| n_k - 1 \right\rangle = \left\langle n_k - 1 \middle| \hat{q}_k \middle| n_k \right\rangle = \sqrt{\frac{\hbar}{2\omega_k} \cdot n_k}$$

$$\left\langle n_k \middle| \hat{p}_k \middle| n_k - 1 \right\rangle = -\left\langle n_k - 1 \middle| \hat{p}_k \middle| n_k \right\rangle = i\sqrt{\frac{\hbar \omega_k}{2} \cdot n_k}$$

(7.3)

Вместо канонических операторов $\hat{q}_k, \hat{p}_k$ удобно использовать их линейные комбинации, имеющие только по одному отличному от ноля матричному элементу для переходов между собственными состояниями гамильтониана (6.4):

$$\hat{c}_k = \frac{1}{\sqrt{2\hbar\omega_k}} \cdot (\omega_k \hat{q}_k + i\hat{p}_k) \qquad\qquad \hat{c}_k^+ = \frac{1}{\sqrt{2\hbar\omega_k}} \cdot (\omega_k \hat{q}_k - i\hat{p}_k) \qquad (7.4)$$

Несложно показать, что

$$\langle n_k - 1|\hat{c}_k|n_k\rangle = \sqrt{n_k} \qquad\qquad \langle n_k|\hat{c}_k^+|n_k - 1\rangle = \sqrt{n_k} \qquad (7.5)$$

Правило коммутации для $\hat{c}_k, \hat{c}_k^+$ принимает вид:

$$\hat{c}_k \hat{c}_k^+ - \hat{c}_k^+ \hat{c}_k = 1 \qquad (7.6)$$

Таким образом мы можем рассматривать операторы $\hat{c}_k, \hat{c}_k^+$ как операторы уничтожения и рождения плазмона.

Операторы $\hat{\vec{E}}^{(u)}, \hat{\vec{H}}^{(u)}, \hat{A}^{(u)\mu}$ могут быть выражены через $\hat{c}_k, \hat{c}_k^+$:

$$\hat{\vec{E}}^{(u)} = \sum_{\vec{k}} \frac{\exp(q^{(u)} \cdot z)}{2\sqrt{\Omega_k}} \left( \hat{c} \cdot \vec{e}^{(u)} \cdot \exp(ik_x \cdot x + ik_y \cdot y - i\omega t) + \hat{c}^+ \cdot \vec{e}^{(u)*} \cdot \exp(-ik_x \cdot x - ik_y \cdot y + i\omega t) \right)$$

$$\hat{\vec{H}}^{(u)} = \sum_{\vec{k}} \frac{\exp(q^{(u)} \cdot z)}{2\sqrt{\Omega_k}} \left( \hat{c} \cdot \vec{h}^{(u)} \cdot \exp(ik_x \cdot x + ik_y \cdot y - i\omega t) + \hat{c}^+ \cdot \vec{h}^{(u)*} \cdot \exp(-ik_x \cdot x - ik_y \cdot y + i\omega t) \right) \qquad (7.7)$$

$$\hat{A}^{(u)\mu} = \sum_{\vec{k}} \frac{\exp(q^{(u)} \cdot z)}{2\sqrt{\Omega_k}} \left( \hat{c} \cdot a^{(u)\mu} \cdot \exp(ik_x \cdot x + ik_y \cdot y - i\omega t) + \hat{c}^+ \cdot a^{(u)\mu*} \cdot \exp(-ik_x \cdot x - ik_y \cdot y + i\omega t) \right)$$

где

$$\Omega_k = \frac{\Theta_k}{2\hbar\omega_k} = \frac{S}{16\pi \cdot \hbar\omega_k} \cdot \frac{k^2 \cdot (\varepsilon_h - \varepsilon_m)}{\varepsilon_h p^2 q} \left( 1 + \frac{qd}{2} \left( \frac{(\varepsilon_m + \varepsilon_h)}{\varepsilon_h} - \frac{\varepsilon_m \omega^2}{k^2} \right) \right) \qquad {\color{red}(7.8)}$$

8. Вычислим вероятность излучения плазмона в единицу времени излучателем, расположенным вблизи тонкой пленки металла. Для определенности будем считать, что излучатель расположен в области выше пленки (см. рис. 1), при $z > z_h$. Размеры излучателя будем считать малыми, сравнительно с длиной волны плазмона, и вероятность вычислим в дипольном приближении. Будем следовать методу расчета вероятности спонтанного излучения, изложенному в /8,§45/ для случая излучателя в свободном пространстве.

Взаимодействие электромагнитного поля с зарядом описывается оператором:

$$\hat{V} = e \cdot \iiint dV \cdot \hat{A}_\mu \hat{j}^\mu \qquad (8.1)$$

Здесь $e$ — заряд электрона, $\hat{A}_\mu, \hat{j}^\mu$ — вторично квантованные операторы потенциала электромагнитного поля и "плотности тока электрона". Плотность тока выражается через оператор волновой функции и матрицы Дирака:

$$\hat{j}^\mu = \hat{\bar{\psi}} \gamma^\mu \hat{\psi} \qquad (8.2)$$

Интегрирование в (8.1) выполняется по всему 3-пространству.

Согласно теории возмущений вероятность перехода в единицу времени системы из состояния, описываемого набором квантовых чисел $i$, в состояние $f$ выражается формулой:

$$w_{fi} = \frac{2\pi}{\hbar} |V_{fi}|^2 \cdot \delta(E_f - E_i - \hbar\omega) \qquad (8.3)$$

Матричный элемент оператора возмущения $V_{fi}$ вычисляется с помощью невозмущенных волновых функций состояний $i$ и $f$. Будем считать для простоты, что как излучатель, так и поле имеют дискретный спектр состояний.

9. Выберем калибровку потенциала $A_\mu$ так, чтобы обращалась в ноль компонента $A_0$. Как легко убедиться, сравнивая с (7.7), в такой калибровке потенциал приобретает вид:

$$\hat{A}_\mu = \sum_k \frac{\exp(-q \cdot (z-d))}{2\sqrt{\Omega_k} \cdot \sqrt{\varepsilon_h}\,\omega}\left(\hat{c} \cdot b_{k,\mu} \cdot \exp(ik_x \cdot x + ik_y \cdot y - i\omega t) + \hat{c}^+ \cdot b^*_{k,\mu} \cdot \exp(-ik_x \cdot x - ik_y \cdot y + i\omega t)\right) \qquad (9.1)$$

где $b_\mu$ — единичный вектор с компонентами (контравариантными) (см. Прил.2):

$$b_k^1 = b_{k,x} = -\frac{k_x}{k}\frac{q}{\sqrt{\varepsilon_h}\,\omega} \qquad b_k^2 = b_{k,y} = -\frac{k_y}{k}\frac{q}{\sqrt{\varepsilon_h}\,\omega} \qquad b_k^3 = b_{k,z} = -\frac{ik}{\sqrt{\varepsilon_h}\,\omega} \qquad b^0 = 0 \qquad (9.2)$$

Тогда (8.1) перепишется в 3-х мерном виде:

$$\hat{V} = -e \cdot \iiint dV \cdot \hat{\vec{A}} \cdot \hat{\vec{j}} \qquad (9.3)$$

Спонтанному излучению плазмона соответствует матричный элемент $\left\langle 1_k \left| \hat{A}_\mu \right| 0_k \right\rangle$:

$$\left\langle 1_k \left| \bar{A}_\mu \right| 0_k \right\rangle = \frac{\exp(-q \cdot (z-d))}{2\sqrt{\Omega_k} \cdot \sqrt{\varepsilon_h}\,\omega} \cdot b^*_{k,\mu} \cdot \exp(-ik_x \cdot x - ik_y \cdot y + i\omega t) \qquad (9.4)$$

В дипольном приближении считают величины $A_\mu(x,y,z)$ медленно меняющимися в пределах характерных размеров излучателя и выносят их из-под знака интеграла. В результате матричный элемент оператора (9.3) приобретает вид:

$$\left\langle 1_k f \left| \hat{V} \right| 0_k i \right\rangle = -e \cdot \left\langle 1_k \left| \vec{A}_\kappa (\vec{r}_{em}) \right| 0_k \right\rangle \cdot \iiint dV \cdot \psi_f^* \gamma^0 \vec{\gamma} \psi_i \qquad (9.5)$$

где через $\vec{r}_{em}$ обозначены координаты точки, где находится излучатель. Интеграл по объему в (9.5) в нерелятивистском приближении есть матричный элемент скорости электрона, вычисляемый по нерелятивистским собственным функциям. Матричный элемент скорости связан с матричным элементом координаты и частотой перехода: $\vec{v}_{fi} = -i\omega \vec{r}_{fi}$. Вводя дипольный момент системы $\vec{d} = e\vec{r}$, находим для квадрата модуля матричного элемента:

$$\left| \left\langle 1_k f \left| \hat{V} \right| 0_k i \right\rangle \right|^2 = \left| \vec{d}_{fi} \cdot \vec{b}_k^* \right|^2 \cdot \frac{\exp(-2q \cdot (z_{em} - d))}{4\Omega_k \cdot \varepsilon_h} \qquad (9.6)$$

10. Будем считать, что излучатели вблизи пленки ориентированы хаотично. Усредним выражение (9.6) по направлению дипольного момента излучателя. Это среднее выражается интегралом, который с помощью (9.2) можно записать в виде:

$$\overline{\left| \vec{d}_{fi} \cdot \vec{b}_k^* \right|^2} = \left| \vec{d}_{fi} \right|^2 \cdot \frac{1}{4\pi} \int \left| \frac{q}{\sqrt{\varepsilon_h}\,\omega} \cdot \sin\vartheta \cdot \cos\varphi - \frac{ik}{\sqrt{\varepsilon_h}\,\omega} \cdot \cos\vartheta \right|^2 do \qquad (10.1)$$

(пусть ось $x$ направлена вдоль вектора $\vec{k}$, так что $b_{k,y} = 0$).

Вычисление интеграла (10.1) приводит к ответу (см. Приложение 3):

$$\overline{\left| \vec{d}_{fi} \cdot \vec{b}_k^* \right|^2} = \left| \vec{d}_{fi} \right|^2 \cdot \frac{1}{3} \cdot \frac{q^2 + k^2}{\varepsilon_h \omega^2} \qquad (10.2)$$

Из (8.3) находим среднее значение вероятности перехода в единицу времени с излучением плазмона с волновым вектором $\vec{k}$:

$$\overline{w_{fi}(k)} = \frac{2\pi}{\hbar} \left| \vec{d}_{fi} \right|^2 \cdot \frac{1}{3} \cdot \frac{q^2 + k^2}{\varepsilon_h \omega^2} \cdot \frac{\exp(-2q \cdot (z_{em} - d))}{4\Omega_k \cdot \varepsilon_h} \cdot \delta(E_f - E_i - \hbar\omega) \qquad (10.3)$$

11. Для вычисления полной вероятности перехода с излучением плазмона просуммируем величину (10.3) по всем модам, определяемым набором $k_x, k_y$. Сумму ряда по дискретному набору состояний заменим интегралом, учитывая при этом, что на каждый интервал значений $\Delta k_x$ приходится $\Delta N_x = \dfrac{\Delta k_x}{(2\pi/a)}$ состояний поля, где $a$ — сторона выбранного прямоугольника $S$. Аналогичное соотношение справедливо для $\Delta k_y$.

Таким образом мы приходим к выражению для вероятности спонтанного перехода:

$$\overline{w_{fi}} = \int_{-\infty}^{\infty} \frac{a \cdot dk_x}{2\pi} \int_{-\infty}^{\infty} \frac{b \cdot dk_y}{2\pi} \frac{2\pi}{\hbar} \left| \vec{d}_{fi} \right|^2 \cdot \frac{1}{3} \cdot \frac{q^2 + k^2}{\varepsilon_h \omega^2} \cdot \frac{\exp(-2q \cdot (z_{em} - d))}{4\Omega_k \cdot \varepsilon_h} \cdot \delta(E_f - E_i - \hbar\omega) \tag{11.1}$$

Перейдем к полярным координатам в плоскости $(k_x, k_y)$ и проинтегрируем по азимутальному углу:

$$\overline{w_{fi}} = \frac{S}{3\hbar} \cdot \left| \vec{d}_{fi} \right|^2 \cdot \int_0^{\infty} k \, dk \cdot \frac{q^2 + k^2}{\varepsilon_h \omega^2} \cdot \frac{\exp(-2q \cdot h)}{4\Omega_k \cdot \varepsilon_h} \cdot \delta(E_f - E_i - \hbar\omega) \tag{11.2}$$

где $h = z_{em} - d$ — расстояние от излучателя до поверхности пленки.

Переходя в (11.2) к интегрированию по $d\omega$ и вычисляя интеграл $\delta$-функции, найдем

$$\overline{w_{fi}} = \frac{S}{3\hbar^2} \cdot \left| \vec{d}_{fi} \right|^2 \cdot k \cdot \left( \frac{\partial k}{\partial \omega} \right) \cdot \frac{q^2 + k^2}{\varepsilon_h \omega^2} \cdot \frac{\exp(-2q \cdot h)}{4\Omega_k \cdot \varepsilon_h} \tag{11.3}$$

Или с учетом (7.8):

$$\overline{w_{fi}} = \frac{4\pi}{3\hbar} \cdot \left| \vec{d}_{fi} \right|^2 \cdot \left( \frac{\partial k}{\partial \omega} \right) \cdot \frac{q^2 + k^2}{k \cdot (\varepsilon_h - \varepsilon_m) \cdot \varepsilon_h \omega} \cdot p^2 q \cdot \exp(-2q \cdot h) \cdot \left( 1 + \frac{qd}{2} \left( \frac{(\varepsilon_m + \varepsilon_h)}{\varepsilon_h} - \frac{\varepsilon_m \omega^2}{k^2} \right) \right)^{-1} \tag{11.4}$$

Сравним полученное выражение с известной формулой вероятности дипольного излучения в свободном пространстве /8,§45/:

$$w_{ph} = \frac{4\omega^3}{3\hbar} \cdot \left| \vec{d}_{fi} \right|^2 \tag{11.5}$$

Введем фактор $F$, характеризующий возрастание вероятности спонтанного излучения плазмона вблизи тонкой пленки, относительно вероятности излучения фотона:

$$F = \frac{\overline{w_{fi}}}{w_{ph}} = \pi \cdot \left( \frac{\partial k}{\partial \omega} \right) \cdot \frac{q^2 + k^2}{k \cdot (\varepsilon_h - \varepsilon_m) \cdot \varepsilon_h \omega^4} \cdot p^2 q \cdot \exp(-2q \cdot h) \cdot \left( 1 + \frac{qd}{2} \left( \frac{(\varepsilon_m + \varepsilon_h)}{\varepsilon_h} - \frac{\varepsilon_m \omega^2}{k^2} \right) \right)^{-1} \tag{11.6}$$

12. Рассчитаем зависимости $\omega_k, \Omega_k, F$ для пленки серебра толщиной $3\,nm$. Не будем учитывать поглощения в серебре, так как оптическом диапазоне частот $\varepsilon''/\varepsilon' \approx 10^{-2}$. Плазменная частота для серебра составляет $\omega_{pl} = 9.1\,eV$. Диэлектрическую проницаемость среды, окружающей пленку примем

$$\varepsilon_h = 1.6 \tag{12.1}$$

На рис. 2 представлен результат расчета частоты поверхностной волны в зависимости от величины волнового вектора $k$. Из рисунка видно, что частотам оптического диапазона соответствуют значения волнового вектора в интервале $0.02\,nm^{-1} < k < 0.14\,nm^{-1}$, что соответствует поверхностным волнам с длиной волны в интервале $40\,nm < \lambda < 300\,nm$.

Построим зависимость от волнового вектора $k$ амплитуды нормированного потенциала $\widehat{A}_\mu$ в непосредственной близости от поверхности пленки (ср. (9.1),(9.2)) :

$$R_k = \sqrt{A_{kz} \cdot A_{kz}^*} = \frac{k}{\sqrt{\Omega_k \cdot \varepsilon_h \cdot \omega^2}}$$

(12.2)

График зависимости $R_k$ представлен на рис. 3.

На рис. 4 представлена зависимость от величины волнового вектора плазмона фактора $F$ (см.(11.6)), характеризующего увеличение вероятности спонтанного излучения атома вблизи тонкой пленки, сравнительно с вероятностью спонтанного перехода в свободном пространстве. Мы видим, что в диапазоне оптических частот фактор $F$ возрастает от величины $3 \cdot 10^2$ на красном крае спектра до величины $10^5$ на фиолетовом крае.

Автор благодарен В.С.Зуеву и А.М.Леонтовичу за многократные и содержательные дискуссии, сыгравшие существенную роль в данной работе.

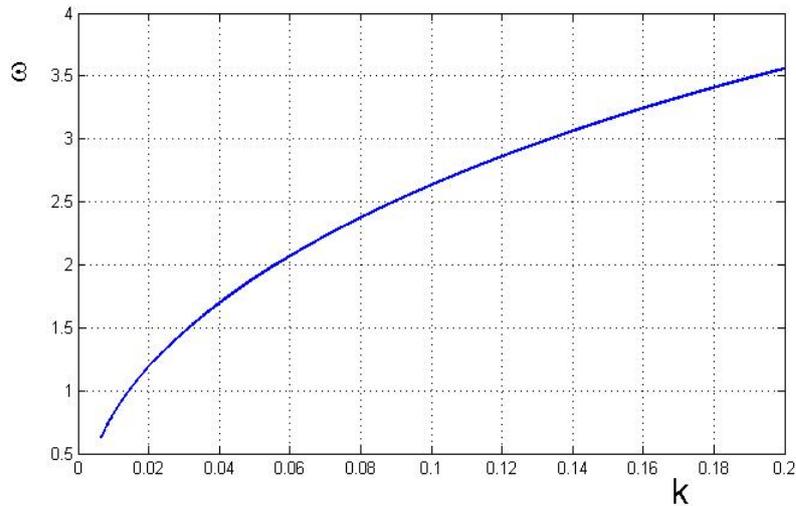

Рис. 2. Зависимость от модуля волнового вектора частоты свободной поверхностной волны в пленке серебра толщины $d = 3\,nm$. По оси абсцисс отложена величина волнового вектора $k, nm^{-1}$, а по оси ординат частота волны $\omega_k, eV$.

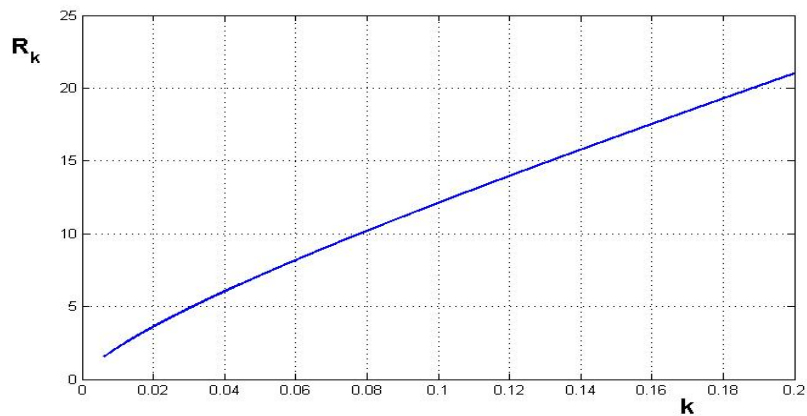

Рис. 3. Зависимость от волнового вектора $k, [nm^{-1}]$ амплитуды $R_k, [a.u.]$ (см.(12.2)) потенциала нормированной волновой функции плазмона . Параметры расчета в тексте.

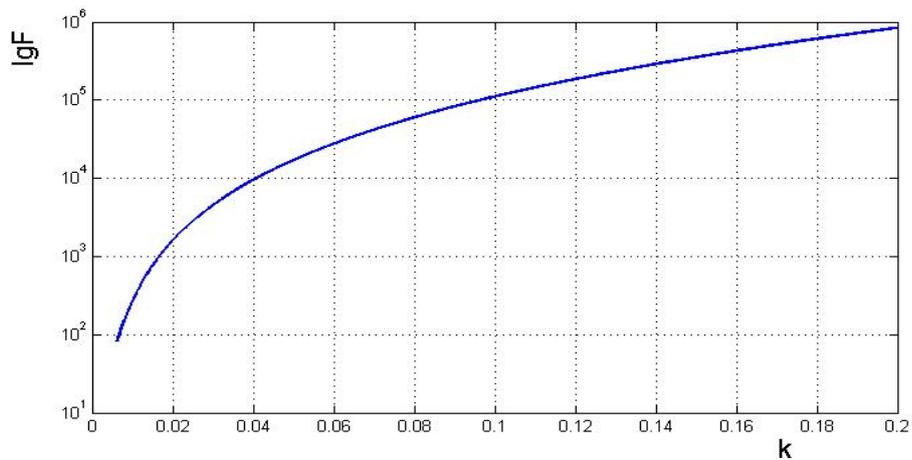

Рис. 4. Зависимость от волнового вектора фактора увеличения вероятности спонтанного излучения плазмона относительно вероятности излучения фотона. Волновой вектор в $\left[nm^{-1}\right]$.

Приложение 1.
Вычисление интеграла (5.6)

$$W_k = \frac{S}{8\pi} \cdot \left( \int\limits_{-\infty}^{0} dz (\varepsilon \vec{E}_k \vec{E}_k^* + \vec{H}_k \vec{H}_k^*) + \int\limits_{0}^{d} dz (\varepsilon \vec{E}_k \vec{E}_k^* + \vec{H}_k \vec{H}_k^*) + \int\limits_{d}^{\infty} dz (\varepsilon \vec{E}_k \vec{E}_k^* + \vec{H}_k \vec{H}_k^*) \right) \quad (5.6)$$

$$\vec{E}_k^{(u)}(t,z) = \gamma \cdot \vec{e}_k^{(u)} \cdot \exp(-i\omega t) \cdot \exp(q^{(u)} z)$$
$$\vec{H}_k^{(u)}(t,z) = \gamma \cdot \vec{h}_k^{(u)} \cdot \exp(-i\omega t) \cdot \exp(q^{(u)} z) \quad (4.6)$$

$$e_{k\xi}^{(h)} = -\frac{iq}{\varepsilon_h \cdot \omega} \cdot \exp(qd) \qquad e_{k\xi}^{(m)} = -\frac{ip}{\varepsilon_m \cdot \omega} \cdot \frac{K+1}{2} \qquad e_{k\xi}^{(r)} = -\frac{ip}{\varepsilon_m \cdot \omega} \cdot \frac{K-1}{2} \qquad e_{k\xi}^{(b)} = -\frac{iq}{\varepsilon_h \cdot \omega}$$

$$e_{kz}^{(h)} = \frac{k}{\varepsilon_h \cdot \omega} \cdot \exp(qd) \qquad e_{kz}^{(m)} = -\frac{k}{\varepsilon_m \cdot \omega} \cdot \frac{K+1}{2} \qquad e_{kz}^{(r)} = \frac{k}{\varepsilon_m \cdot \omega} \cdot \frac{K-1}{2} \qquad e_{kz}^{(b)} = \frac{k}{\varepsilon_h \cdot \omega} \quad (4.8)$$

$$h_{k\eta}^{(h)} = -\exp(qd) \qquad h_{k\eta}^{(m)} = \frac{K+1}{2} \qquad h_{k\eta}^{(r)} = -\frac{K-1}{2} \qquad h_{k\eta}^{(b)} = 1$$

$$I_1 = \int\limits_{-\infty}^{0} dz (\varepsilon \vec{E}_k \vec{E}_k^* + \vec{H}_k \vec{H}_k^*) = \gamma\gamma^* \int\limits_{-\infty}^{0} \exp(2qz) dz \cdot (\varepsilon_h e_{k\xi}^{(b)} e_{k\xi}^{(b)*} + \varepsilon_h e_{kz}^{(b)} e_{kz}^{(b)*} + h_{k\eta}^{(b)} h_{k\eta}^{(b)*})$$

$$I_1 = \gamma\gamma^* \frac{1}{2q} \cdot \left( \varepsilon_h \cdot \frac{q^2}{\varepsilon_h^2 \omega^2} + \varepsilon_h \cdot \frac{k^2}{\varepsilon_h^2 \omega^2} + 1 \right) = \gamma\gamma^* \frac{1}{2q} \left( \frac{q^2 + k^2 + \varepsilon_h \omega^2}{\varepsilon_h \omega^2} \right) = \gamma\gamma^* \frac{k^2}{q \cdot \varepsilon_h \omega^2}$$

$$I_3 = \int\limits_{d}^{\infty} dz (\varepsilon \vec{E}_k \vec{E}_k^* + \vec{H}_k \vec{H}_k^*) = \gamma\gamma^* \int\limits_{d}^{\infty} \exp(-2qz) dz \cdot (\varepsilon_h e_{k\xi}^{(h)} e_{k\xi}^{(h)*} + \varepsilon_h e_{kz}^{(h)} e_{kz}^{(h)*} + h_{k\eta}^{(h)} h_{k\eta}^{(h)*})$$

$$I_3 = \gamma\gamma^* \frac{\exp(-2qd)}{2q} \cdot \left( \varepsilon_h \cdot \frac{q^2}{\varepsilon_h^2 \omega^2} + \varepsilon_h \cdot \frac{k^2}{\varepsilon_h^2 \omega^2} + 1 \right) \cdot \exp(2qd) = \gamma\gamma^* \frac{1}{2q} \cdot \left( \frac{q^2 + k^2 + \varepsilon_h \omega^2}{\varepsilon_h \omega^2} \right) = \gamma\gamma^* \frac{k^2}{q \cdot \varepsilon_h \omega^2}$$

$$I_2 = \int\limits_{0}^{d} dz (\varepsilon \vec{E}_k \vec{E}_k^* + \vec{H}_k \vec{H}_k^*)$$

$$I_2 = \gamma\gamma^* \int\limits_{0}^{d} e^{2pz} dz \cdot (\varepsilon_m e_{k\xi}^{(m)} e_{k\xi}^{(m)*} + \varepsilon_m e_{kz}^{(m)} e_{kz}^{(m)*} + h_{k\eta}^{(m)} h_{k\eta}^{(m)*}) + \gamma\gamma^* \int\limits_{0}^{d} e^{-2pz} dz \cdot (\varepsilon_m e_{k\xi}^{(r)} e_{k\xi}^{(r)*} + \varepsilon_m e_{kz}^{(r)} e_{kz}^{(r)*} + h_{k\eta}^{(r)} h_{k\eta}^{(r)*})$$

$$+ \gamma\gamma^* \int\limits_{0}^{d} dz \cdot (\varepsilon_m e_{k\xi}^{(m)} e_{k\xi}^{(r)*} + \varepsilon_m e_{k\xi}^{(m)*} e_{k\xi}^{(r)} + \varepsilon_m e_{kz}^{(m)} e_{kz}^{(r)*} + \varepsilon_m e_{kz}^{(m)*} e_{kz}^{(r)} + h_{k\eta}^{(m)} h_{k\eta}^{(r)*} + h_{k\eta}^{(m)*} h_{k\eta}^{(r)})$$

$$I_2 = \gamma\gamma^* \cdot \frac{e^{2pd} - 1}{2p} \cdot \left( \varepsilon_m \cdot \frac{p^2}{\varepsilon_m^2 \omega^2} + \varepsilon_m \cdot \frac{k^2}{\varepsilon_m^2 \omega^2} + 1 \right) \cdot \left( \frac{K+1}{2} \right)^2$$

$$\gamma\gamma^* \cdot \frac{1 - e^{-2pd}}{2p} \cdot \left( \varepsilon_m \cdot \frac{p^2}{\varepsilon_m^2 \omega^2} + \varepsilon_m \cdot \frac{k^2}{\varepsilon_m^2 \omega^2} + 1 \right) \cdot \left( \frac{K-1}{2} \right)^2$$

$$+ \gamma\gamma^* \cdot d \cdot \left( 2 \cdot \varepsilon_m \cdot \frac{p^2}{\varepsilon_m^2 \omega^2} - 2 \cdot \varepsilon_m \cdot \frac{k^2}{\varepsilon_m^2 \omega^2} - 2 \right) \cdot \left( \frac{K^2 - 1}{4} \right)$$

$$I_2 = \gamma\gamma^* \cdot \frac{e^{2pd}-1}{2p} \cdot 2 \cdot \frac{k^2}{\varepsilon_m \omega^2} \cdot \left(\frac{K+1}{2}\right)^2 + \gamma\gamma^* \cdot \frac{1-e^{-2pd}}{2p} \cdot 2 \cdot \frac{k^2}{\varepsilon_m \omega^2} \cdot \left(\frac{K-1}{2}\right)^2$$
$$+ \gamma\gamma^* \cdot 2 \cdot d \cdot \left(\frac{p^2-k^2}{\varepsilon_m \omega^2}-1\right) \cdot \left(\frac{K^2-1}{4}\right)$$

$$I_2 = \gamma\gamma^* \cdot \frac{k^2}{4 \cdot p \cdot \varepsilon_m \omega^2} \cdot \left((e^{2pd}-1) \cdot (K+1)^2 + (1-e^{-2pd}) \cdot (K-1)^2\right)$$
$$- \gamma\gamma^* \cdot 4 \cdot d \cdot \left(\frac{K^2-1}{4}\right)$$

$$I_2 = \gamma\gamma^* \cdot \frac{k^2}{4 \cdot p \cdot \varepsilon_m \omega^2} \cdot \left(e^{2pd} \cdot (K+1)^2 - e^{-2pd} \cdot (K-1)^2 + (K-1)^2 - (K+1)^2\right)$$
$$- \gamma\gamma^* \cdot d \cdot (K^2-1)$$

Учтем (2.5) и (4.5)

$$\exp(p \cdot d) = \frac{(\varepsilon_m \cdot q - \varepsilon_h \cdot p)}{(\varepsilon_m \cdot q + \varepsilon_h \cdot p)}$$

$$K = \frac{\varepsilon_m \cdot q}{\varepsilon_h \cdot p}$$

$$\exp(p \cdot d) = \frac{(K-1)}{(K+1)}$$

$$I_2 = \gamma\gamma^* \cdot \frac{k^2}{4 \cdot p \cdot \varepsilon_m \omega^2} \cdot \left((K-1)^2 - (K+1)^2 + (K-1)^2 - (K+1)^2\right) - \gamma\gamma^* \cdot d \cdot (K^2-1)$$

$$I_2 = \gamma\gamma^* \cdot \frac{k^2}{2 \cdot p \cdot \varepsilon_m \omega^2} \cdot \left((K-1)^2 - (K+1)^2\right) - \gamma\gamma^* \cdot d \cdot (K^2-1)$$

$$I_2 = -\gamma\gamma^* \cdot \frac{2 \cdot k^2}{p \cdot \varepsilon_m \omega^2} \cdot K - \gamma\gamma^* \cdot d \cdot (K^2-1)$$

$$I_2 = -\gamma\gamma^* \cdot \frac{2 \cdot k^2 \cdot q}{p^2 \cdot \varepsilon_h \omega^2} - \gamma\gamma^* \cdot d \cdot \left(\left(\frac{\varepsilon_m \cdot q}{\varepsilon_h \cdot p}\right)^2 - 1\right)$$

$$I_1 + I_2 + I_3 = -\gamma\gamma^* \cdot \frac{2 \cdot k^2 \cdot q}{p^2 \cdot \varepsilon_h \omega^2} - \gamma\gamma^* \cdot d \cdot \left(\left(\frac{\varepsilon_m \cdot q}{\varepsilon_h \cdot p}\right)^2 - 1\right) + 2 \cdot \gamma\gamma^* \frac{k^2}{q \cdot \varepsilon_h \omega^2}$$

$$I_1 + I_2 + I_3 = -\gamma\gamma^* \cdot d \cdot \left(\frac{\varepsilon_m^2 \cdot q^2 - \varepsilon_h^2 \cdot p^2}{\varepsilon_h^2 \cdot p^2}\right) + 2 \cdot \gamma\gamma^* \frac{k^2}{q \cdot \varepsilon_h \omega^2 \cdot p^2} \cdot (p^2 - q^2)$$

$$I_1 + I_2 + I_3 = -\gamma\gamma^* \cdot d \cdot \left(\frac{\varepsilon_m^2 \cdot (k^2 - \varepsilon_h \omega^2) - \varepsilon_h^2 \cdot (k^2 - \varepsilon_m \omega^2)}{\varepsilon_h^2 \cdot p^2}\right) + 2 \cdot \gamma\gamma^* \frac{k^2}{q \cdot \varepsilon_h \omega^2 \cdot p^2} \cdot (-\varepsilon_m \omega^2 + \varepsilon_h \omega^2)$$

$$I_1 + I_2 + I_3 = -\gamma\gamma^* \cdot d \cdot \left(\frac{(\varepsilon_m^2 - \varepsilon_h^2) \cdot k^2 - \varepsilon_h \varepsilon_m \omega^2 \cdot (\varepsilon_m - \varepsilon_h)}{\varepsilon_h^2 \cdot p^2}\right) + 2 \cdot \gamma\gamma^* \frac{k^2}{q \cdot \varepsilon_h \cdot p^2} \cdot (-\varepsilon_m + \varepsilon_h)$$

$$I_1 + I_2 + I_3 = \gamma\gamma^* \cdot (\varepsilon_h - \varepsilon_m) \cdot \left(d \cdot \left(\frac{(\varepsilon_m + \varepsilon_h) \cdot k^2 - \varepsilon_h \varepsilon_m \omega^2}{\varepsilon_h^2 \cdot p^2}\right) + 2 \cdot \frac{k^2}{q \cdot \varepsilon_h \cdot p^2}\right)$$

$$I_1 + I_2 + I_3 = \gamma\gamma^* \cdot (\varepsilon_h - \varepsilon_m) \cdot \frac{2 \cdot k^2}{q \cdot \varepsilon_h \cdot p^2} \cdot \left(1 + \frac{qd}{2} \cdot \left(\frac{(\varepsilon_m + \varepsilon_h) \cdot k^2 - \varepsilon_h \varepsilon_m \omega^2}{\varepsilon_h \cdot k^2}\right)\right)$$

$$I_1 + I_2 + I_3 = \gamma\gamma^* \cdot (\varepsilon_h - \varepsilon_m) \cdot \frac{2 \cdot k^2}{q \cdot \varepsilon_h \cdot p^2} \cdot \left(1 + \frac{qd}{2} \cdot \left(\frac{(\varepsilon_m + \varepsilon_h)}{\varepsilon_h} - \frac{\varepsilon_m \omega^2}{k^2}\right)\right)$$

Приложение 2.

$$\vec{E} = -\partial_t \vec{A} - \nabla\Phi \quad \vec{H} = rot\vec{A}$$

Калибровочное преобразование:

$$\vec{A}' = \vec{A} + \nabla\chi \,; \quad \Phi' = \Phi - \partial_t \chi$$

$$\chi = \frac{a^{(h)0}}{-i\omega} \cdot \exp(-qz - i\omega t + ik_x x + ik_y y)$$

$$\Phi' = 0$$

$$b^1 = \frac{a^{(h)0}}{-i\omega} \cdot ik_x = -\frac{k_x \cdot q}{k \cdot \varepsilon_h \omega^2} \exp(qd)$$

$$b^2 = \frac{a^{(h)0}}{-i\omega} \cdot ik_y = -\frac{k_y \cdot q}{k \cdot \varepsilon_h \omega^2} \exp(qd)$$

$$b^3 = -\frac{i}{k} \cdot \exp(qd) - \frac{a^{(h)0}}{-i\omega} \cdot q = -\frac{i}{k} \cdot \exp(qd) - \frac{iq^2}{k \cdot \varepsilon_h \omega^2} \exp(qd)$$

$$b^3 = -\frac{i}{k} \cdot \exp(qd) \cdot \left(1 + \frac{q^2}{\varepsilon_h \omega^2}\right) = -\frac{ik}{\varepsilon_h \omega^2} \cdot \exp(qd)$$

$$\vec{b}\vec{b} = k_x^2 \cdot \left(\frac{q}{k \cdot \varepsilon_h \omega^2} \exp(qd)\right)^2 + k_y^2 \cdot \left(\frac{q}{k \cdot \varepsilon_h \omega^2} \exp(qd)\right)^2 + \left(\frac{ik}{\varepsilon_h \omega^2} \exp(qd)\right)^2$$

$$\vec{b}\vec{b} = \left(\frac{q}{\varepsilon_h \omega^2} \exp(qd)\right)^2 + \left(\frac{ik}{\varepsilon_h \omega^2} \exp(qd)\right)^2$$

$$\vec{b}\vec{b} = \left(\frac{1}{\varepsilon_h \omega^2} \exp(qd)\right)^2 \cdot (q^2 - k^2) = -\frac{\exp(2qd)}{\varepsilon_h \omega^2}$$

Приложение 3.

$$\left|\vec{d}_{fi} \cdot \vec{b}_k^*\right|^2 = \left|\vec{d}_{fi}\right|^2 \cdot \frac{1}{4\pi} \int\left(\frac{q}{\sqrt{\varepsilon_h}\omega} \cdot \sin\vartheta \cdot \cos\varphi - \frac{ik}{\sqrt{\varepsilon_h}\omega} \cdot \cos\vartheta\right)\left(\frac{q}{\sqrt{\varepsilon_h}\omega} \cdot \sin\vartheta \cdot \cos\varphi + \frac{ik}{\sqrt{\varepsilon_h}\omega} \cdot \cos\vartheta\right)do$$

$$\left|\vec{d}_{fi} \cdot \vec{b}_k^*\right|^2 = \left|\vec{d}_{fi}\right|^2 \cdot \frac{1}{4\pi} \int\left(\frac{q^2}{\varepsilon_h \omega^2} \cdot \sin^2\vartheta \cdot \cos^2\varphi + \frac{k^2}{\varepsilon_h \omega^2} \cdot \cos^2\vartheta\right) \cdot \sin\vartheta d\vartheta d\varphi$$

$$\left|\vec{d}_{fi} \cdot \vec{b}_k^*\right|^2 = \left|\vec{d}_{fi}\right|^2 \cdot \frac{1}{2} \int\left(\frac{q^2}{2\varepsilon_h \omega^2} \cdot \sin^2\vartheta + \frac{k^2}{\varepsilon_h \omega^2} \cdot \cos^2\vartheta\right) \cdot \sin\vartheta d\vartheta$$

$$\left|\vec{d}_{fi} \cdot \vec{b}_k^*\right|^2 = \left|\vec{d}_{fi}\right|^2 \cdot \frac{1}{2} \int_{-1}^{1} \left( \frac{q^2}{2\varepsilon_h \omega^2} \cdot (1-u^2) + \frac{k^2}{\varepsilon_h \omega^2} \cdot u^2 \right) du$$

$$\left|\vec{d}_{fi} \cdot \vec{b}_k^*\right|^2 = \left|\vec{d}_{fi}\right|^2 \cdot \frac{1}{2} \left( \frac{q^2}{2\varepsilon_h \omega^2} \cdot \frac{4}{3} + \frac{k^2}{\varepsilon_h \omega^2} \cdot \frac{2}{3} \right)$$

$$\left|\vec{d}_{fi} \cdot \vec{b}_k^*\right|^2 = \left|\vec{d}_{fi}\right|^2 \cdot \frac{1}{3} \cdot \frac{q^2 + k^2}{\varepsilon_h \omega^2}$$